\documentclass[sigconf, authorversion]{acmart}

\usepackage{tikz}
\usepackage{pgfplots}
\usepackage{pgfplotstable}
\usepackage{subcaption}
\usepackage{multirow}
\usepackage{multicol}
\usepackage{graphicx}
\usepackage{array}
\usepackage{float}

\AtBeginDocument{%
  \providecommand\BibTeX{{%
    \normalfont B\kern-0.5em{\scshape i\kern-0.25em b}\kern-0.8em\TeX}}}

\copyrightyear{2024}
\acmYear{2024}
\setcopyright{rightsretained}
\acmConference[ITiCSE 2024]{Proceedings of the 2024 Conference on Innovation and Technology in Computer Science Education V. 1}{July 8--10, 2024}{Milan, Italy}
\acmBooktitle{Proceedings of the 2024 Conference on Innovation and Technology in Computer Science Education V. 1 (ITiCSE 2024), July 8--10, 2024, Milan, Italy}\acmDOI{10.1145/XXXXXX.XXXXXX}
\acmISBN{979-8-4007-XXXX-X/XX/XX}

\begin{document}

\title[Explaining Code with a Purpose]{Explaining Code with a Purpose: An Integrated Approach for Developing Code Comprehension and Prompting Skills}

\author{Paul Denny}
\orcid{0000-0002-5150-9806}
\affiliation{
  \institution{The University of Auckland}
  \city{Auckland}
  \country{New Zealand}
}
\email{paul@cs.auckland.ac.nz}

\author{David H. Smith IV}
\orcid{0000-0002-6572-4347}
\affiliation{
  \institution{University of Illinois}
  \city{Urbana, IL}
  \country{USA}
}
\email{dhsmith2@illinois.edu}

\author{Max Fowler}
\orcid{0000-0002-4730-447X}
\affiliation{
  \institution{University of Illinois}
  \city{Urbana, IL}
  \country{USA}
}
\email{mfowler5@illinois.edu}

\author{James Prather}
\orcid{0000-0003-2807-6042}
\affiliation{
  \institution{Abilene Christian University}
  \city{Abilene, TX}
  \country{USA}
}
\email{james.prather@acu.edu}

\author{Brett A. Becker}
\orcid{0000-0003-1446-647X}
\affiliation{
  \institution{University College Dublin}
  \city{Dublin}
  \country{Ireland}
}
\email{brett.becker@ucd.ie}

\author{Juho Leinonen}
\orcid{0000-0001-6829-9449}
\affiliation{
  \institution{University of Auckland}
  \city{Auckland}
  \country{New Zealand}
}
\email{juho.2.leinonen@aalto.fi}

\renewcommand{\shortauthors}{Paul Denny, et al.}

\begin{abstract}
Reading, understanding and explaining code have traditionally been important skills for novices learning programming.  As large language models (LLMs) become prevalent, these foundational skills are more important than ever given the increasing need to understand and evaluate model-generated code.  Brand new skills are also needed, such as the ability to formulate clear prompts that can elicit intended code from an LLM.  Thus, there is great interest in integrating pedagogical approaches for the development of both traditional coding competencies and the novel skills required to interact with LLMs.  One effective way to develop and assess code comprehension ability is with ``Explain in plain English'' (EiPE) questions, where students succinctly explain the purpose of a fragment of code.  However, grading EiPE questions has always been difficult given the subjective nature of evaluating written explanations and this has stifled their uptake.  In this paper, we explore a natural synergy between EiPE questions and code-generating LLMs to overcome this limitation.  We propose using an LLM to generate code based on students' responses to EiPE questions -- not only enabling EiPE responses to be assessed automatically, but helping students develop essential code comprehension and prompt crafting skills in parallel.  We investigate this idea in an introductory programming course and report student success in creating effective prompts for solving EiPE questions.  We also examine student perceptions of this activity and how it influences their views on the use of LLMs for aiding and assessing learning.
\end{abstract}



\maketitle

\begin{figure}[h!]
\centering
  \includegraphics[width=\linewidth]{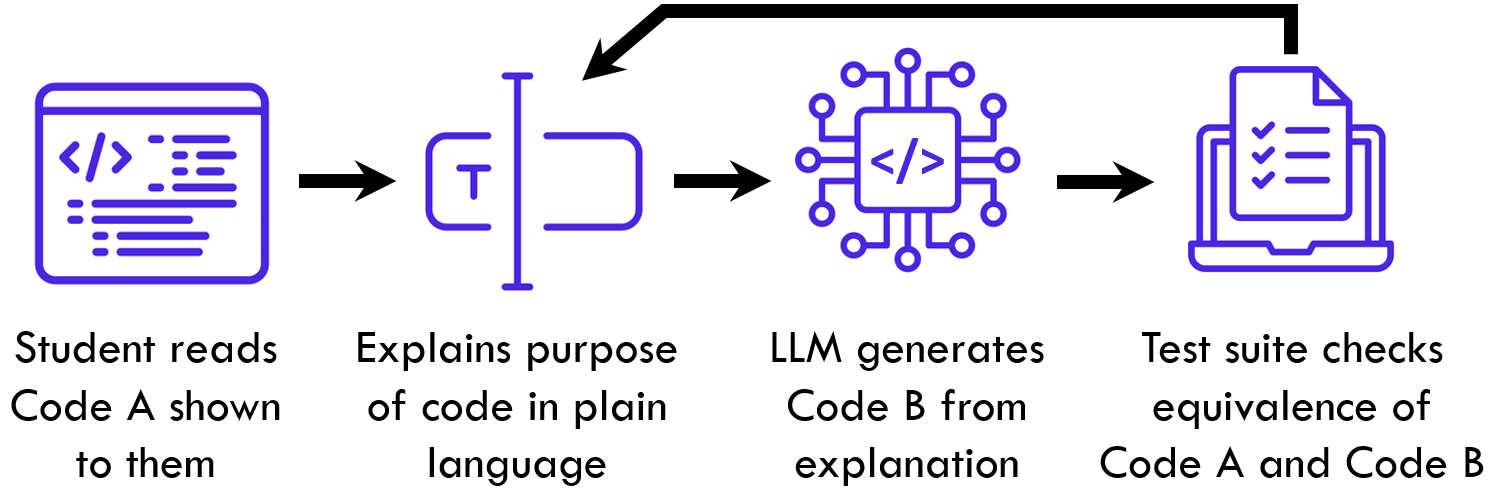}
  \caption{Explaining the purpose of a code fragment to an LLM until it generates functionally equivalent code, targeting code comprehension and prompt crafting skills in parallel.}  \label{fig:schematic}
\end{figure}

\section{Introduction}

The ability to read and comprehend source code is an essential skill for all
programmers, from novices to professionals \cite{medvidova2022program}. Indeed,
due to the ease of generating code with large language models (LLMs),
programmers will be spending an increasing proportion of their time
understanding and evaluating LLM-generated code \cite{becker2023programming}. Of
course, teaching students to comprehend code is not a new challenge; reading and
tracing code execution and answering comprehension questions about code have
been common and effective strategies in programming courses for many years
\cite{swidan2019effect, busjahn2011analysis, hassan2021exploring,
sudol2012code}.  In particular, EiPE (``Explain in Plain English'') questions
are a widely studied format for assessing how well students can read and
understand code at an abstract level \cite{murphy2012explain,
corney2014explain}.

Manual grading approaches for EiPE questions have been informed by the
``Structure of Observed Learning Outcome'' (SOLO) taxonomy where a given
response can be categorized based on the degree to which it integrates all
elements of a given segment of code in order to describe that code's purpose
rather than its implementation~\cite{biggs2014evaluating, lister2006not}. As a
response relating the various elements of a topic together is treated as one of
the highest levels of comprehension in the SOLO taxonomy, a typical EiPE
question asks students to describe the \emph{purpose} of a provided piece of
code in natural language~\cite{murphy2012ability}. Through interviews with
members of the computing education research community, Fowler et
al.~\cite{fowler2021how} found that EiPE questions were valued for their
abstraction aspect which was seen to aid in debugging, communication and
functional decomposition.  Despite this clear potential, the difficulty of
grading EiPE questions -- due to the subjective nature of evaluating students'
written explanations -- has been a longstanding barrier to their wide-scale
adoption \cite{li2023wrong, weeda2020towards, fowler2021autograding}.
As a remedy to this shortcoming, \citet{smith2023code}
proposed an automated EiPE grading approach that involves generating code from
a student's EiPE response to determine if their description
is accurate.

In the current paper, we explore this potential synergy between EiPE questions and
code-generating LLMs. Illustrated in Figure \ref{fig:schematic}, our method
begins by presenting a student with a fragment of code, much like in a
traditional EiPE question. The student then reads and attempts to understand
what the code is doing, and crafts an explanation of the code in natural
language.  To assess this explanation, we use the method
proposed by \citet{smith2023code} where the student's response is provided as
the input prompt to a code-generating LLM. The code that is generated is
automatically tested for equivalence with the original code using a test suite.
This approach enables the objective evaluation of responses to EiPE questions
and supports the development of both code comprehension skills and skills
related to clearly formulating prompts for LLMs.

To investigate the potential of this idea, we deployed a series of these
problems to students in a large introductory programming course (n$\approx$900).
First, we examined how well students solved these tasks, by  analysing success
rates and prompt lengths (with and without a character limit enforced) and we
classified prompts with respect to the SOLO taxonomy.  Second, we investigated
students' perceptions of this activity which was their first experience using an
LLM to assess code comprehension skill.  Students compared the activity to more
familiar programming tasks, and indicated the extent to which they felt it was a
valid way to evaluate their learning.  We organise our study, and the
presentation of results, around the following three research questions:

\vspace{1mm}

\begin{enumerate}
    \item[\textbf{RQ1:}] How successful are students at crafting code explanation prompts that induce an LLM to generate functionally equivalent code?
    \item[\textbf{RQ2:}] What is the relationship, if any, between the success of a student's prompt and the classification of that prompt with respect to the categories of the SOLO taxonomy?  
    \item[\textbf{RQ3:}] What are students' thoughts about the code explanation activity in comparison to more traditional code writing tasks, and do they see it as an accurate way to assess their comprehension of code? 
\end{enumerate}

\section{Related work}

\subsection{Large Language Models in Introductory Programming Courses}

The advent of large language models (LLMs) is ushering in significant changes to
the landscape of programming education \cite{denny2024computing}.  A range of novel 
tools and pedagogical approaches are being developed and 
evaluated, including the use of digital teaching assistants and automated resource generation ~\cite{sheese2024patterns, sarsa2022automatic}, and commercial tools like GitHub
Copilot have been seamlessly integrated into widely-used Integrated Development
Environments (IDEs).  This has begun to revolutionize the code-writing process
for both seasoned developers and new learners. From a computing education
perspective, the ubiquity and capability of LLMs has raised questions regarding
academic integrity and how and when such tools should be integrated into CS1
courses~\cite{prather2023robots, sheard2024instructor,
raman2022programming}.

Studies investigating the capabilities of these models, namely Codex, GPT-3, and
GPT-4, have shown that they can solve typical programming problems at least as
well as an average introductory programming student~\cite{finnie2022robots,
finnieansley2023my, cipriano2023gpt}. Recent work by Denny et al. evaluated the
ability of the Codex model to solve introductory programming exercises from
natural language specifications \cite{denny2023conversing}. Of the 166 problems
they evaluated, approximately 50\% could be solved by simply supplying the
original question prompt, and 80\% could be solved after small manual
modifications were made to the prompts. Though this may raise concerns related
to the ease with which students are able to generate solutions, it also
highlights how a human can work iteratively with an LLM to refine a prompt to
generate desired code.  This suggests that explicit instruction on how to
formulate effective prompts, supported by appropriate practice opportunities,
could be beneficial for students learning programming.

\subsection{Explain in Plain English Questions \& Prompt Problems}

The issue of problem formulation when prompting for code generation bears some
resemblance to ``Explain in Plain English'' (EiPE) questions. Both require
students to formulate a description of some code. In the case of EiPE questions,
``correctness'' of a response is generally evaluated based on whether it
unambiguously conveys the functionality of the code at a
high-level~\cite{azad2020strategies, fowler2021should}. This approach to grading
has been largely inspired by the SOLO taxonomy as it has been applied to code
comprehension.  This differentiates between a student describing the structures
present in a segment of code from it's higher level purpose. The latter is
considered to demonstrate a higher level of comprehension and is more typical of
descriptions produced by experts~\cite{lister2006not}. On the other hand,
successful prompting might be characterized by the ability to provide a
description that elicits code that functions as the prompter intended.

Beyond the relationship between EiPE questions and prompting there is an
emerging notion that prompting, evaluating the resulting code, and then
potentially re-prompting is an emerging skillset which CS students should be
explicitly taught~\cite{denny2024computing}. Early work in this direction
includes the research around ``Prompt problems'' \cite{denny2024promptproblems}, 
where students are shown images that illustrate how
inputs should be transformed to outputs and are tasked with constructing
solutions in natural language. Similarly, \citet{smith2023code} evaluated a
grading approach for EiPE questions which they term ``Code Generation Based
Grading'' (CGBG). This involved collecting students' responses to EiPE questions
taken from a large historical dataset, and grading them based on the correctness
of the code they produce when used as a prompt to an LLM.    They note that
future work should investigate the utility of this grading approach as a tool
for teaching prompting skills, given that effectively utilizing the system's
feedback requires students to evaluate the relationship between their prompt,
the code it produced, and the code they were attempting to describe.

\section{Methods}

We conducted our study in an introductory programming course at the University
of Auckland, a large public research University in New Zealand.  The 12-week
course is required for all students in the Engineering program, and introduces
standard CS1 topics using a combination of MATLAB and C.  In the spring term of
2023, when our data was collected, 889 students were enrolled.

\subsection{Code Explanation Tasks}

Throughout the course, students complete weekly laboratory exercises primarily
consisting of sets of programming tasks that are automatically graded. For the
purposes of the current study, we included the new code explanation tasks in two
lab sessions.  For convenience in the paper, we will refer to these lab sessions
as Lab A and Lab B, although they were held in Week 2 and Week 4 of the C
programming module respectively.  The topic of Lab A was ``Loops, Arrays and
Functions'' and the topic of Lab B was ``Strings,  Text Processing and 2D
arrays''.  The new tasks were delivered using a modified version of the
open-source PrairieLearn platform~\cite{west2015prairielearn}.  Alongside
regular programming tasks, both Lab A and Lab B included four new code
explanation tasks. Every task consisted of a single function (named `foo') that
had to be explained -- Lab A did not enforce any length limit on the
explanations that students wrote, whereas for Lab B a 250-character limit was
imposed.  

\begin{figure}
\centering
  \includegraphics[width=.9\linewidth]{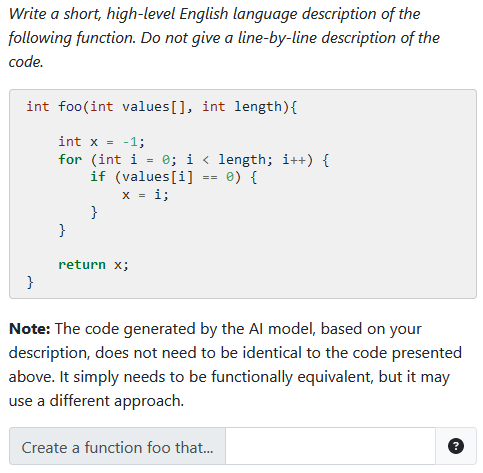}
  \caption{A problem presented in the form of a function implementation with obfuscated identifier names; students enter a natural language description that will be used to prompt an LLM to generate equivalent code to that shown.}
  \vspace{-3mm}
  \label{fig:pl-indexlastzero}
\end{figure}

A short description of each task is listed in Table
\ref{tab:correctness_attempts} when presenting the results (e.g. ``Index of last
zero'').  Figure \ref{fig:pl-indexlastzero} shows a screenshot of the
PrairieLearn platform with one of the code explanation tasks from Lab A. The
task shown is the ``Index of last zero'' task, where the code returns the index
position of the rightmost occurrence of the value 0, or -1 if it is not present.
Students complete the prompt, which begins \emph{``Create a function foo
that...''}, and submit it for grading. Students were also given the following
information about the task: 

\emph{``The goal of this task is for you to read and understand the purpose of
the code shown below. Describe the purpose of the code using plain language;
your description will be given to an AI language model, and the model will
generate code matching your description. You will have solved the task when the
code generated by the AI model is functionally equivalent to the code you have
described.''} Upon submitting a prompt, the code generated by the LLM is
displayed along with the results of the test cases.  

The tasks were graded, but contributed only a small fraction towards each
student's final score (approximately 1\%).  No penalties were given for
incorrect submissions, but for each task a maximum of 20 attempts were allowed.

\subsection{Analysis of Students' Prompts} 

In addressing \textbf{RQ1} and \textbf{RQ2} we operationalize success at
completing the prompting task as the ability to provide a prompt that (1) is
successful in generating code that passes the provided test cases and (2)
demonstrates comprehension of the code's purpose. To evaluate the latter, we use
the ``Structures of Observed Learning Outcome'' (SOLO)
taxonomy~\cite{biggs2014evaluating}. Specifically, we use the adapted SOLO
taxonomy presented by \citet{lister2006not} as well as their process for
applying these codes to student's responses. The taxonomy and definitions we use
when coding students' prompts are:
\begin{itemize}
    \item \textbf{Prestructural:} A student demonstrates one or more significant misconceptions or no understanding of the code.
    \item \textbf{Unistructural:} A student demonstrates some understanding of the code or the code's purpose but provides an incomplete description or the description contains some misconceptions.
    \item \textbf{Multistructural} The student provides a correct and complete description of the code and its structures but does not fully join these descriptions together to describe the code's overall purpose.
    \item \textbf{Relational:} The student demonstrates a correct and high-level understanding of the code's purpose by relating all of its elements together and describing the code's purpose.
\end{itemize}
We also include a fifth category which we term \textbf{Direct Recitation}. This
category includes responses where a student directly recited the code in a
line-by-line fashion or directly copied the code or elements of the code
verbatim into the prompt without demonstrating any understanding of the code's
structures. 

Student prompts were categorized by two members of the research team. From each
of the eight questions, 200 prompts were randomly selected for deductive coding
using the categories described above. In the event that a student's prompt
contained a \textit{multistructural} description in addition to a
\textit{relational} summary, that response was graded as \textit{relational}.
For the purposes of establishing inter-rater reliability (IRR) the response set
from two questions were coded independently by each of the researchers.
Inter-rater reliability was calculated using Cohen's kappa and found to be 0.79,
well above the accepted threshold for high IRR~\cite{mchugh2012interrater}. The
researchers then met to reconcile those disagreements that did exist. Given the
high IRR, the researchers then each coded 100 prompts from each of the remaining
six questions independently for a total of 1600 prompts.

\begin{table}[t]
\centering
\small
\begin{tabular}{|l|p{6cm}|}
\hline
\textbf{Type} & \textbf{Question} \\ \hline
\begin{tabular}[t]{@{}l@{}}Open \\ (prompted)\end{tabular} & Please reflect on the code comprehension tasks and comment on what you think about them compared to typical programming tasks. \\ \hline
\begin{tabular}[t]{@{}l@{}}Likert \\ (SD,D,N,A,SA)\end{tabular} 
& Having AI language models generate code from natural language descriptions is an accurate way to evaluate code comprehension skills. \\ \hline
\begin{tabular}[t]{@{}l@{}}Open \\ (unprompted)\end{tabular} & Do you have any comments about this lab? \\ \hline
\end{tabular}
\caption{Reflection questions related to RQ3.}\label{tab:studentperceptionquestions}
\vspace{-7mm}
\end{table}

\subsection{Student Perceptions}
To address \textbf{RQ3}, after completing (or attempting) the code explanation
tasks students were asked to reflect on the activity by responding to three
questions.  Two of these were open-response questions, and the other used a
standard 5-point Likert scale.  The first open-response question directly asked
students to comment on the code explanation task, and thus we consider this
`prompted' feedback.  The other open-response question was a generic question
about any aspect of the lab, and thus we consider any comments relating to the
code explanation task in response to this question to be `unprompted'.  The
three questions are listed in Table \ref{tab:studentperceptionquestions}. We
summarize responses to the Likert item using a diverging stacked bar chart. We
analyzed open-response data using the guidelines for reflexive thematic analysis
outlined by Braun and Clarke \cite{braun2006using}.  This includes a coding
phase in which responses are tagged with succinct labels, followed by phases of
generating and developing higher-level themes collated from these
labels\footnote{\url{https://www.thematicanalysis.net/doing-reflexive-ta/}}.
When presenting the results in Section \ref{sec:RQ3}, we report the most common
themes and illustrate these with examples of student responses.

\section{Results}

\subsection{RQ1: Task Completion Success}

Overall, students experienced a high degree of success in both Lab A and Lab B,
with almost all students completing all tasks
(Table~\ref{tab:correctness_attempts}). Despite being given 20 attempts per
task, the majority of students completed each of the tasks using only one or two
attempts.

\begin{figure}
 \centering
 \includegraphics[width=.95\columnwidth]{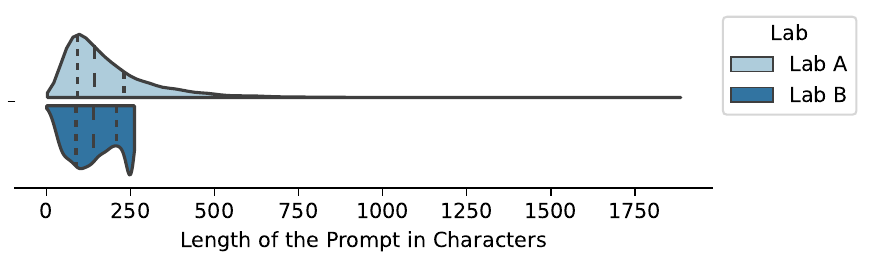}
 \vspace{-4mm}
 \caption{Distribution of prompt lengths for each of the two lab sessions (Lab B enforced a 255 character limit).}
 \vspace{-4mm}
 \label{fig:prompt_length_dists}
\end{figure}

Lab B differed from Lab A in that it imposed a 255 character limit on students'
prompts. Examining the distribution of prompt lengths for each of the two labs
reveals that the median response length for each of the two labs was similar
(Figure~\ref{fig:prompt_length_dists}). The primary impact of including the
character limit appears to be that it eliminated the small number of extremely
verbose prompts seen in Lab A. The inclusion of the character limit does not
appear to have had a significant impact on students' success at the task as the
average number of attempts for each of the labs is quite similar
(Table~\ref{tab:correctness_attempts}).

\begin{table}[]
\small
\begin{tabular}{|l|l|c|c|c|}
\hline
\# &\textbf{Task Description} & \textbf{$\mu$} & \textbf{$\sigma$} & \textbf{\% Correct} \\ \hline
\multirow{4}{*}{\rotatebox[origin=c]{90}{Lab A  \hspace{0.03cm}}} & Sum between a and b inclusive & 2.04          & 2.53             & 98.2               \\ \cline{2-5} 
                               & Count even numbers in array & 1.37          & 1.00              & 99.0               \\ \cline{2-5} 
                               & Index of last zero  & 2.18 & 3.44 & 99.8 \\ \cline{2-5} 
                               & Sum positive values& 1.43          & 0.75             & 99.2               \\ \hline
\multirow{4}{*}{\rotatebox[origin=c]{90}{Lab B \hspace{0.2cm}}} & Reverse a string                                  & 1.65          & 1.56             & 99.6               \\ \cline{2-5} 
                               & Calculate sum of row in 2D array  & 2.03 & 3.11 & 98.1 \\ \cline{2-5} 
                               & Is a vowel contained in a string?                     & 1.49          & 1.89             & 98.1               \\ \cline{2-5} 
                               & Does a string contain a substring? & 2.58 & 6.43 & 96.3 \\ \hline
\end{tabular}
\caption{The number of submission attempts and percentage of students who successfully completed each task.}\label{tab:correctness_attempts}
\vspace{-5mm}
\end{table}

\subsection{RQ2: SOLO Category \& Prompt Success}

\begin{figure}
    \centering
    \includegraphics[width=.95\columnwidth]{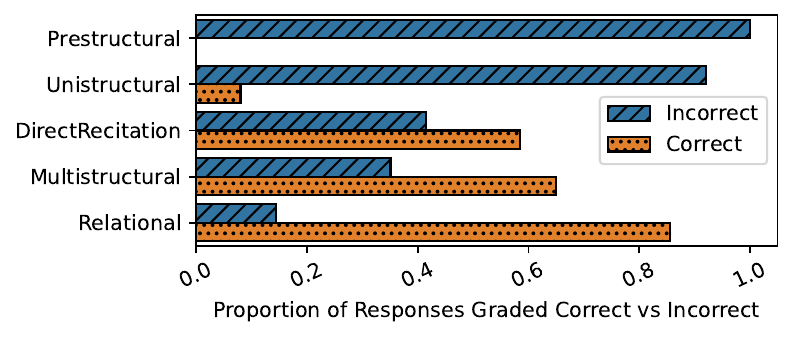}
    \vspace{-4mm}
    \caption{Proportion of prompts generating correct and incorrect code at each SOLO level.}
    \label{fig:correct_v_incorrect}
\end{figure}

\begin{figure}
    \centering
    \includegraphics[width=.95\columnwidth]{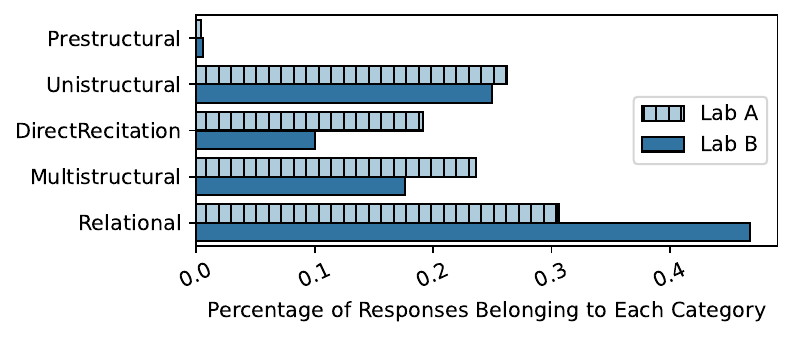}
    \vspace{-4mm}
    \caption{Proportion of prompts classified at each SOLO-level for each lab.}
    \vspace{-4mm}
    \label{fig:a_v_b}
\end{figure}

From Figure~\ref{fig:correct_v_incorrect}, we see that prompts that contained
small or significant misconceptions (\textit{Unistructural} and
\textit{Prestructural}) generated code which was near universally graded as
incorrect. Though this might be expected for prestructural it is positive that
the LLM, in general, did not generate correct code for a prompt that contained a
misconception. If it had done so, then this approach to grading EiPE
questions could reinforce misconceptions.

Prompts classified as \textit{Direct Recitation} and \textit{Multistructural}
more often generated code which was graded as correct rather than incorrect,
but not overwhelmingly so. It may be the case that such prompts can contain
certain ambiguities or lack specific details which human graders are willing
to overlook in the event the prompt contains the core ideas the grader is
looking for. However, these ambiguities may lead an LLM to generating plausible
though ultimately incorrect interpretations of a student's prompt.

Finally, \textit{Relational} prompts generated code which was overwhelmingly
graded as correct. This finding is encouraging as it indicates a synergy between
successful prompting, which focuses on successfully generating code, and high
quality EiPE responses, which focus on high-level descriptions of code. 

In comparing the results for Lab A and Lab B, we found that Lab B included a
higher proportion of relational responses than Lab A (Figure~\ref{fig:a_v_b}).
It may be the case that the character limit imposed for Lab B not only reduced
the presence of extremely long explanations but increased the presence of
relational ones as well. However, there are alternative explanations.  Students’
code comprehension skills may have improved between the two labs, and students’
may have found relational responses more successful in Lab A and thus
deliberately aimed to construct them in Lab B. In any case, given the success of
relational prompts for generating code, future work seeking to teach students
how to successfully prompt LLMs should further explore approaches to encouraging
relational prompts.

\begin{table*}[htbp]
\centering
\footnotesize
\begin{tabular}{|l|p{12.5cm}|p{1.9cm}|}
\hline
\textbf{Task description} & \textbf{Longest correct response} & \textbf{Shortest correct response} \\ \hline

\begin{tabular}[t]{@{}l@{}}Count even numbers \\ in array\end{tabular} & takes an array of values called 'values{[}{]}' and the length (or number of values contained within the array) of the array. The program then uses a for loop with the conditionals i initialized to equal zero, looping the for loop for as long as i is less than the length of the array and adding 1 (using i++) every iteration. This is essentially scans the length of the array of values and operates on each item in the array. The operation for each item in the array (i.e. values{[}i{]}, the operation on the number at any given index) is to test wether the value leaves any remainder when divided by two (i.e. using mod 2). If the number leaves no remainder when divided by two it is even and the count for x is incremented by one. This continues until every value in the array has been tested. The returned value 'x' is the count of how many numbers left no remainder when divided by two, which is a count of how many even numbers were in the array of values. & find how many numbers divisible by 2 \\ \hline

\begin{tabular}[t]{@{}l@{}}Reverse a string\end{tabular} & The provided code is a C function named `foo` that reverses the characters in a given character array (string) `str`. Here's a breakdown of how the code works: 1. `\#include \textless{}string.h\textgreater{}`: This line includes the standard C library header file `string.h`, wh & flips a string \\ \hline

\begin{tabular}[t]{@{}l@{}}Does a string \\ contain a substring \end{tabular} & The code defines a C function named foo that checks if the second string str2 is contained within the first string str1. It iterates through str1 and compares substrings of the same length as str2. If a match is found, it returns 1; otherwise, it returns & checks if str2 is a in str1 \\ \hline

\end{tabular}
\vspace{2.5mm}
\caption{The longest and shortest correct prompts submitted for a selection of the exercises (the last two, from Lab B, had a 255 character limit enforced).  Some prompts show evidence of being generated by LLMs, given tell-tale language such as ``Here's a breakdown of what happens'' and the start of a list of line-by-line explanations (e.g., the longest response in the middle row).}
 \vspace{-5mm}
\label{tab:longestandshortestprompts}
\end{table*}

\subsection{RQ3: Student Perceptions}
\label{sec:RQ3}

A total of 812 non-empty responses were submitted to the first open-response
question, where students were directly `prompted' on their perceptions of the
code explanation exercises in comparison to traditional code writing tasks.
Thematic analysis of these responses revealed several key themes. Generally
speaking, students enjoyed the exercises and found them interesting, and took a
positive view of their educational value.

\subsubsection{Novelty and Engagement}
The most prominent theme that emerged related to the novelty and engaging nature
of the activity. Students frequently expressed that the task was \emph{``very
interesting and fun as it's nice seeing AI included''} and that it provides a
\emph{``nice change of pace from writing code''}. This sentiment is reflected
clearly in the following responses:

\begin{quote}
\emph{``I actually found it a little bit entertaining since it was unbelievable to me that a code could be written directly from a sentence or two.''}
\end{quote}

\begin{quote}
\emph{``The task was a lot of fun to do and it was interesting to see how the ai understood my work based on my statements.''}
\end{quote}

Many similar comments were observed, indicating that students generally enjoyed
the activity, found it novel, and appreciated that some aspects of AI were
integrated into the course.  

\subsubsection{Enhanced Comprehension of Code}

The next most prevalent theme related to the perception that the activity
enhanced students' comprehension of code. Students often remarked on the
usefulness of the exercise to articulate their understanding in \emph{``plain
English,''} which helped \emph{``improve my skills of understanding code''}:

\begin{quote}
\emph{``I think it is very useful in really comprehending chunks of code in terms of their purpose rather than its individual tasks.''}
\end{quote}

\begin{quote}
\emph{``It actually gets people to read and understand code, and try to figure out what the original code was supposed to do. I feel that my ability to describe the actual action not just the steps going on improved as I read the code more.''}
\end{quote}

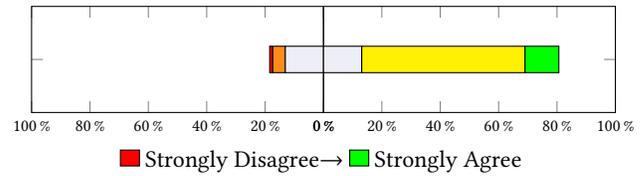
\begin{figure}
    \centering
        \begin{tikzpicture}
    \pgfplotstableread{
Semester  TooMany    Many       Neutral     Few        TooFew
Q6        1.05       4.30       26.16       55.93      11.63
    }\frequency
    \begin{axis}[
        scale only axis,
        name=ax1,
        legend cell align=center,
        legend columns=-1,
        legend style={at={(1,-0.25)},anchor=north,draw=none},
        xbar stacked,
        xmin=-100,
        xmax=0,
        try min ticks=3,
        xticklabel style = {font=\scriptsize},
        xticklabel=\pgfmathparse{abs(\tick)}\pgfmathprintnumber{\pgfmathresult}\,$\%$,
        ytick=data,
        yticklabel style={align=center},
        yticklabels={
        },
        enlarge y limits={abs=0.45cm},
        width=110px,
        height=40px
    ]
        \addlegendimage{fill=red}
        \addlegendimage{fill=green}
        \addplot [fill=gray!70!blue!10] table [x expr=-(\thisrow{Neutral}/2), meta=Semester ,y expr=\coordindex] {\frequency};
        \addplot [fill=orange!90] table [x expr=-\thisrow{Many}, meta=Semester ,y expr=\coordindex] {\frequency};
        \addplot [fill=red] table [x expr=-\thisrow{TooMany}, meta=Semester ,y expr=\coordindex] {\frequency};
        \addlegendentry{Strongly Disagree\textrightarrow} 
        \addlegendentry{Strongly Agree}
    \end{axis}

    \begin{axis}[
        scale only axis,
        at=(ax1.south east),
        xbar stacked,
        xmin=0,
        xmax=100,
        xticklabel style = {font=\scriptsize},
        try min ticks=3,
        xticklabel=\pgfmathparse{abs(\tick)}\pgfmathprintnumber{\pgfmathresult}\,$\%$,
        ytick=data,
        yticklabels={},
        enlarge y limits={abs=0.45cm},
        width=110px,
        height=40px
    ]

        \addplot [fill=gray!70!blue!10] table [x expr=(\thisrow{Neutral}/2), meta=Semester ,y expr=\coordindex] {\frequency};
        \addplot [fill=yellow] table [x=Few, meta=Semester ,y expr=\coordindex] {\frequency};
        \addplot [fill=green] table [x=TooFew, meta=Semester ,y expr=\coordindex] {\frequency};

    \end{axis}
\end{tikzpicture} 
     \hfill
     \vspace{-5mm}
     \caption{Student perceptions of whether the task is an accurate way to evaluate code comprehension skills.}\label{fig:likert-results}
     \vspace{-3mm}
\end{figure}

This theme suggests that translating code to natural language prompts encouraged
deeper cognitive processing of programming concepts and improved understanding,
highlighting the educational value of the task. Moreover, student responses to
the Likert item (summarized in Figure \ref{fig:likert-results}) illustrated
generally strong agreement that using an LLM to generate code from their
explanations is an accurate way to evaluate code comprehension skills.  

\subsubsection{Concerns About Effectiveness for Practice and Assessment}
Not all views about the task were positive.  One relatively common theme exposed
concerns regarding the effectiveness of the task, especially how well it  would
scale to problems where the code is more complex.   In addition, some mostly
positive comments were qualified by skeptical statements about the value of the
task in comparison to more traditional activities, or at least a desire to see a
balance between the use of AI-supported and traditional tasks:

\begin{quote}
\emph{``Pretty cool. The code was simplistic enough for an AI to generate a valid result - but I don't think this would work in more complex problems.''}
\end{quote}

\begin{quote}
\emph{``The task was fairly helpful in understanding what code means. However, I feel like it only has limited use in helping me to write code more easily and efficiently.''}
\end{quote}

\begin{quote}
\emph{``So, I think I mixture of the AI model question and regular coding question will be good, but more regular coding question than the other one.''}
\end{quote}

\subsubsection{General feedback}

The second of the two open-response questions, which asked for general feedback
on any aspect of the lab, elicited a total of 80 `unprompted' responses that
made some mention of the code explanation activity.  The vast majority of these
responses were positive, and most related to a general theme around ``Enjoyment
or fun'', aligning with the most common theme observed in responses to the
`prompted' question around the tasks.  Many students highlighted the novelty of
the activity.  For instance, one student reflected, \emph{``The lab was super
fun, especially PrairieLearn as it was a completely new experience''}.  Others
examples included \emph{``I just want to say I loved the PrairieLearn tasks''}
and \emph{``The PraireLearn part was also very fun to do and use''}.

Another positive theme that was common related to ``Learning and
Understanding'', where students appreciated that the task aided their
comprehension of code. One student commented, \emph{``PrairieLearn gave me much
better insight into the connection between natural language and coding, as small
changes in wording had a large impact on the code''}.  Others noted the benefits
for developing new skills, with one even referring to the skill of `writing code
in plain language': \emph{``I would've liked to use prairielearn bit more and
continue to improve my skills of writing code in plain language''}.

Negative feedback was much less common, and included comments around technical
issues and interface design such as wanting to continue experimenting with the
tool even after successfully solving a problem in order to \emph{``push the
ai''}, as well as a small number of students reporting either the tasks were too
easy or too difficult.

\vspace{-1pt}
\section{Discussion}

When the capabilities of large language models (LLMs) became apparent several
years ago, there was initially widespread concern in academia regarding their
potential misuse by students.  Now, as their presence becomes ubiquitous, there
are calls to explore the integration of LLMs into teaching practice and to use
them to power novel educational tools \cite{denny2024computing}.  This is
especially relevant in the field of computing education, where the dominant
pedagogy has involved frequent and repeated practice at \emph{writing} code
\cite{allen2019analysis}.  The ease with which LLMs can now be used to generate
code from natural language prompts suggests a need to re-evaluate teaching
strategies that focus on the mechanics of syntax and code
writing~\cite{savelka2023thrilled, finnieansley2023my}.  Indeed, a recent global
survey of computing educators revealed a strong expectation that students will
need to be taught how to use generative AI tools, and yet concrete pedagogical
approaches are only just beginning to emerge \cite{prather2023robots}.  

Though these exercises may not be considered traditional EiPE questions given
that they lack explicit checks for many of the requirements present in those
rubrics~\cite{chen2020validated, fowler2021how}, they do represent a similar
form of code comprehension task. From our results we see there appears to exist
overlap between what makes a successful EiPE response and prompts which
successfully generate code. This is echoed in the qualitative results where
students not only found the tasks engaging but often mentioned being engaged in
the process of reading code and attempting to comprehend its purpose. These
findings are promising in that they suggest the task was successful in engaging
students with the process of code comprehension. Furthermore, their success at 
the tasks appears to be related to the level at which they were able to express
that comprehension.

As the name suggests, an implicit requirement for traditional ``Explain in Plain
English'' questions is that the descriptions be provided in English.  This could
place students with poorer English language skills -- but equally good code
comprehension skills -- at a disadvantage.  Indeed, this is the motivation for
``refute'' questions, proposed by Kumar and Raman \cite{kumar2023helping}, which
allow students to demonstrate their comprehension of code but without requiring
English language skill. The powerful language translation capabilities of modern
LLMs means that code explanations can be provided in a wide variety of
languages. We observed several students submitting accurate descriptions of code
in languages other than English. For example, a total of six submissions that
successfully solved the tasks were made in Chinese. Although still fairly
infrequent in our data, this suggests that the current approach could be used to
provide an equitable assessment option in diverse classrooms, and this would be
an interesting direction to further explore. 

Our initial findings suggest several
avenues for future work. First, grading is not entirely reliable as the same
prompt may produce correct code on some occasions and incorrect code on others.
A modified version of the activity could generate multiple completions from the
prompt, to assess how reliable the prompt is.  In addition, a poor ``Explain in
Plain English'' prompt, like a line by line explanation, might still produce the
correct code thus reinforcing that behavior. 
For example, Table \ref{tab:longestandshortestprompts} illustrates that prompts leading to correct solutions often varied greatly in length.
Given the success of
relational prompts in this activity, and their alignment with the goals of
teaching students code comprehension, future work should focus on additional
measures which can be taken to nudge students towards providing relational
prompts.

\section{Conclusion}
As the computing education community continues to grapple with questions around
the integration of large language models (LLMs) into the classroom, in this work
we offer some insights into their potential role for developing both code
comprehension and prompting skills.  We propose an approach where code
comprehension is assessed through the use of ``Explain in Plain English'' (EiPE)
questions, by passing student explanations of code to an LLM for evaluation.  In
an empirical study in a large introductory classroom, we observe high rates of
success for students attempting these kinds of tasks, and find that
higher-level, relational descriptions of code are much more likely to succeed.
Feedback from students indicates that they felt the activity not only helps in
improving understanding of code, but was also novel and highly engaging.  Our
work demonstrates just one possible way that LLMs could be integrated into
programming classrooms, and highlights the need for continued work in this
direction.

\bibliographystyle{ACM-Reference-Format}
\bibliography{iticse}

\end{document}